\documentclass[prb,aps,twocolumn,amsmath,amssymb,showpacs,superscriptaddress,letterpaper]{revtex4}
\usepackage{graphicx}
\usepackage{subfigure}
 
\begin{document}

%
%

\title{Spin-Polarized Semiconductor Induced by Magnetic Impurities in Graphene}

\author{Maria Daghofer}
\email{M.Daghofer@ifw-dresden.de} 
\affiliation{Department of Physics and Astronomy, University of Tennessee,
Knoxville, TN 37996-1200 and
\\Materials Science and Technology Division, ORNL, Oak Ridge,
TN 37831-6032, USA}
\affiliation{IFW Dresden, P.O. Box 27 01 16, D-01171 Dresden, Germany}

\author{Nan Zheng}

\affiliation{Department of Physics and Astronomy, University of Tennessee,
Knoxville, TN 37996-1200 and
\\Materials Science and Technology Division, ORNL, Oak Ridge,
TN 37831-6032, USA}
\affiliation{Department of Computer Science, The College of William
  and Mary, Williamsburg, VA 23187, USA} 

\author{Adriana Moreo}
 
\affiliation{Department of Physics and Astronomy, University of Tennessee,
Knoxville, TN 37996-1200 and
\\Materials Science and Technology Division, ORNL, Oak Ridge,
TN 37831-6032, USA}

\date{\today}

\begin{abstract}
The effective magnetic coupling  between magnetic impurities adsorbed
on graphene, which is mediated by the itinerant graphene electrons,
and its impact on the electrons' spectral density are
studied. The magnetic interaction breaks the symmetry between the
sublattices, leading to  antiferromagnetic order, and a
gap for the itinerant electrons develops. Random doping produces a
semiconductor, but if all or most of the impurities are localized in
the same sublattice the spin degeneracy can be lifted  and a
spin-polarized semiconductor induced.  
\end{abstract}
 

 
\maketitle

\section{Introduction and Model}\label{sec:intro}

Graphene has been the focus of active research in the last few years
due to its peculiar transport
properties~\cite{Novo1,Novo2,Twor,Moro,RMPneto} and its potential
technological applications. For spintronic devices,\cite{Zut}
where the high electron mobility
and tunable electron filling make graphene an interesting
candidate, it is important to investigate magnetism and the
possibility of spin-polarized conduction electrons. Moreover, it would
be desirable to transform graphene from a semi-metal into a
semiconductor. In the present paper, we are going to show that
localized magnetic impurities (MIs) provide a way to open a gap. 

Density Functional
Theory (DFT) calculations suggest that
decorating graphene-nanoribbons with metal ions can lead to opposite spin
polarization in the valence and conduction
bands.\cite{cervantes-sodi:165427} This property is known as {\it
  spin-polarized half-semiconductivity} and it is very desirable for
spintronics  
devices.\cite{prigodin:2002} DFT also predicts that interfaces
between Ni or Co layers and graphene or graphite could induce spin
polarization,\cite{karpan:195419} because only the majority spin
bands of the Co or Ni have an overlap with the graphene bands.
We will show that MIs, which do not directly donate
spin-polarized electrons to the graphene layer, can nevertheless
polarize its electrons in the direction
\emph{opposite} to their own spin. 

\begin{figure}[thbp]
\begin{center}
\subfigure{\includegraphics[width=0.13\textwidth]{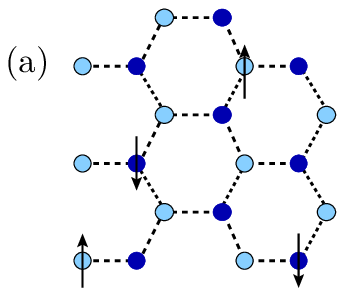}\label{fig:both}}\hspace{0.05\textwidth}
\subfigure{\includegraphics[width=0.13\textwidth]{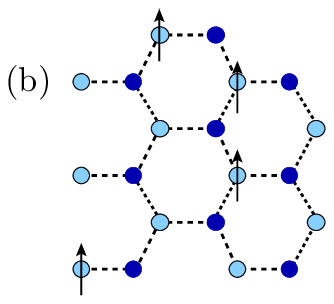}\label{fig:same}}
\caption{(Color online) Cartoon of MIs on graphene. Light (dark)
  circles indicate C atoms in
  sublattice $A$ ($B$). Lines show the NN
  connections, where the electrons hop. Arrows illustrate MIs
  located on top of C atoms, which are either (a) distributed equally
  over both sublattices and AF 
  ordered, or (b) located only in sublattice $A$ and 
  FM ordered.\label{fig:cartoon}}
\end{center}
\end{figure}

Magnetic impurities in graphene can either arise as effective magnetic
moments of non-magnetic
impurities\cite{Kumazaki,kondo_exp}
or from intrinsically MIs. The feasibility of
introducing MIs by embedding transition metal
ions has been studied by {\it ab-initio}
methods~\cite{Krasheninnikov:2009,PhysRevB.81.125433} and can be achieved using a scanning tunneling 
microscope.\cite{uchoa:2008}  Recently, Kondo physics indicating strong coupling between
conduction electrons and MIs has been
reported,\cite{kondo_exp} suggesting that defects indeed
offer a route to tune the electronic properties of graphene.
Single MIs lead to spin-polarized mid-gap states~\cite{lda_defects} and have been
suggested to act as a spin valve.\cite{PhysRevLett.102.046801} The
magnetic (RKKY) coupling between impurities is expected to be
ferromagnetic (FM) [\emph{anti-}ferromagnetic (AF)] for impurities in
the same [different]
sublattices.\cite{Brey,Yaz,Sar,Kumazaki,rkky_ed,PhysRevB.81.125433}
We are going to discuss the ordering of MIs mediated by the
electrons and the spectral properties of these electrons, which then
move in a background of ordered MIs. 

The simplest description of graphene is a tight-binding
Hamiltonian given by the hopping of the carbon-$\pi_z$ electrons
between nearest neighbor (NN) sites of a honeycomb lattice  
(see Fig.~\ref{fig:cartoon}). 
The symmetry between the two C atoms in the unit cell 
leads to the Dirac cones making up the Fermi surface (FS), 
located at the points $K_1$ and $K_2$ [see inset in
Fig.~\ref{fig:bands_both}].  
If this symmetry is broken, e.g., by an energy difference between the
sublattices, a gap can be opened, and 
dispersive edge states within the gap were proposed to arise
at domain boundaries~\cite{graphene_imp_sublatt} or
edges~\cite{Yao:2009p2427} of areas with broken sublattice symmetry. 
The sublattices could be made inequivalent by a substrate with a matching lattice
constant~\cite{Giovannetti:2007p2420} or via non-magnetic impurities, but only if they mostly
occupy sites in only one sublattice.\cite{graphene_imp_sublatt} 
As it will be shown in this letter, the situation is different - and
possibly simpler - for MIs. Due to their AF coupling, they turn out 
to be oriented in opposite directions in the two sublattices and 
an electron that gains magnetic energy in one sublattice, looses it
in the other. Thus, the sublattices are inequivalent for either spin
projection, and both are expected to develop a gap regardless of the
impurity distribution. If, however, MIs or a 
magnetic substrate can be engineered to interact predominantly with
only one sublattice, electrons at the top (bottom) of the valence
(conduction) band will be shown to be spin polarized.


The itinerant electrons interacting with  randomly located MIs 
-- represented by an onsite magnetic moment ${\bf
  S_I}$~\cite{foot1}\nocite{Brey,Yaz,Voz,Duffy} -- are described by
the Hamiltonian 
\begin{equation}\label{eq:ham}
H=-t \sum_{{\bf \langle i,j\rangle},\sigma}
(c^{\dagger}_{{\bf i},\sigma}c_{{\bf j},\sigma}+\text{h.c.})
-{2J\sum_{{\bf I}}{\bf s_I}\cdot{\bf S_I}}\;,
\end{equation}
where $c^{\dagger}_{{\bf i},\sigma}$ creates an electron with spin
$\sigma$,
${\bf \langle i,j \rangle}$ are NN sites in 
the honeycomb lattice, 
$J$ gives the strength of the
magnetic interaction between the impurities and the electrons, and the spin 
of the electrons is given 
by ${\bf s_I}=\sum_{\alpha,\beta}c^{\dagger}_{{\bf I},\alpha}
\sigma_{\alpha,\beta}c_{{\bf I},\beta}$. 
The system is half-filled, i.e.,
one electron per site, and $t=3\;\textrm{eV}$.\cite{Dress,Voz}
To simplify the calculations and to avoid the `sign problem' in the Monte Carlo (MC) simulations, we treat the MI spins
as classical vectors. This approximation is valid for large impurity
spins $S\gtrsim 1$, as is expected to hold for transition metal
ions adsorbed on graphene,\cite{Duffy} and the localized spins then
simply act as a spin-dependent impurity potential on the itinerant electrons. 
The magnitude of the spin is absorbed into the coupling constant $J$,
giving $2J \approx 3\;\textrm{eV}$.\cite{baskaran} 
Coulomb repulsion between the conduction electrons is neglected, 
as has been found appropriate in many physical
situations.\cite{Brey,Sarma}
Since our lattices are considerably larger than the distances between
impurities, finite-size effects are not expected to be severe.\cite{rkky_ed}

\section{Results and Discussion}\label{sec:res}

\begin{figure}
\subfigure{
\includegraphics[width = 0.2\textwidth]{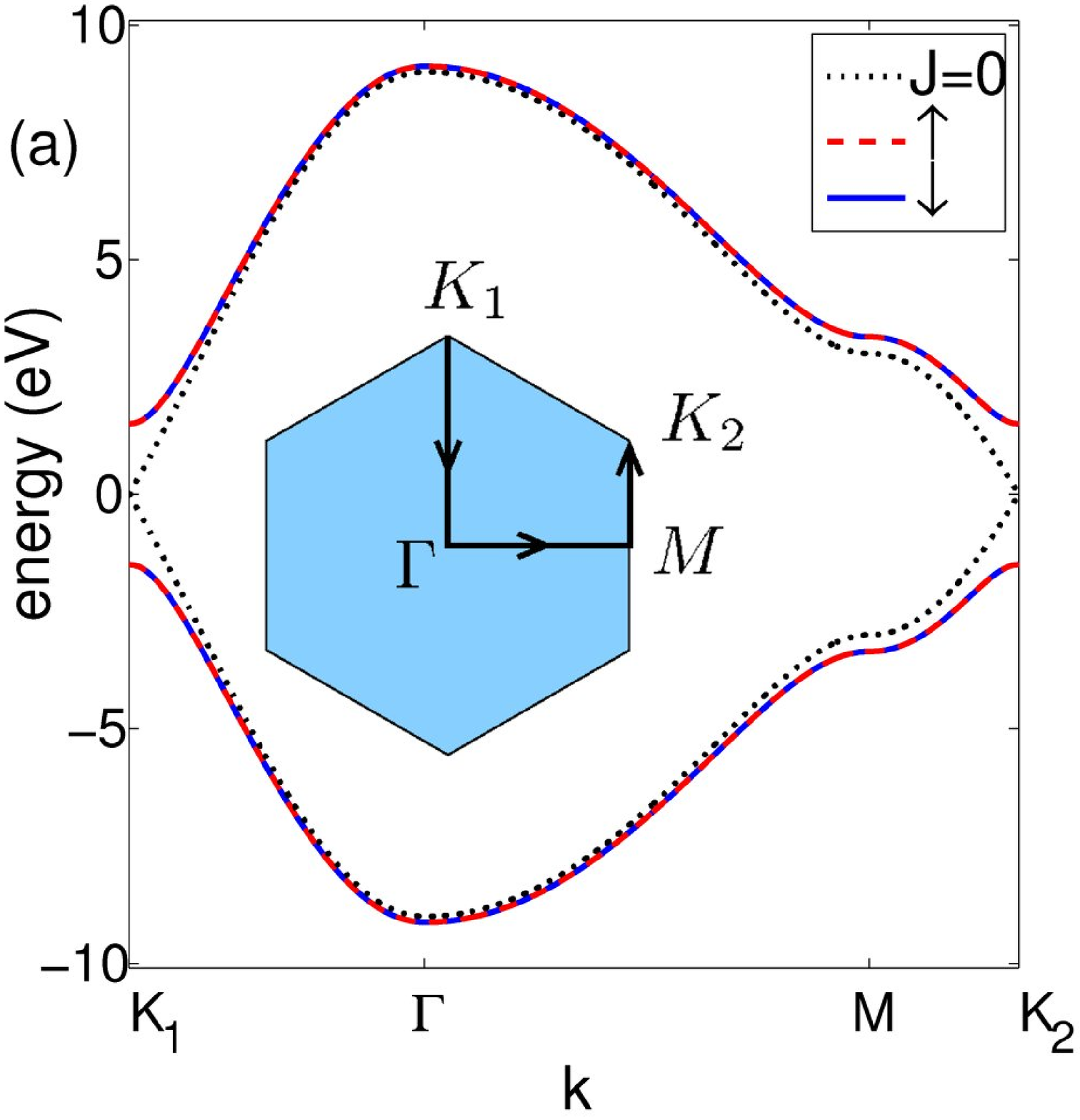}\label{fig:bands_both}}
\subfigure{
\includegraphics[width = 0.2\textwidth]{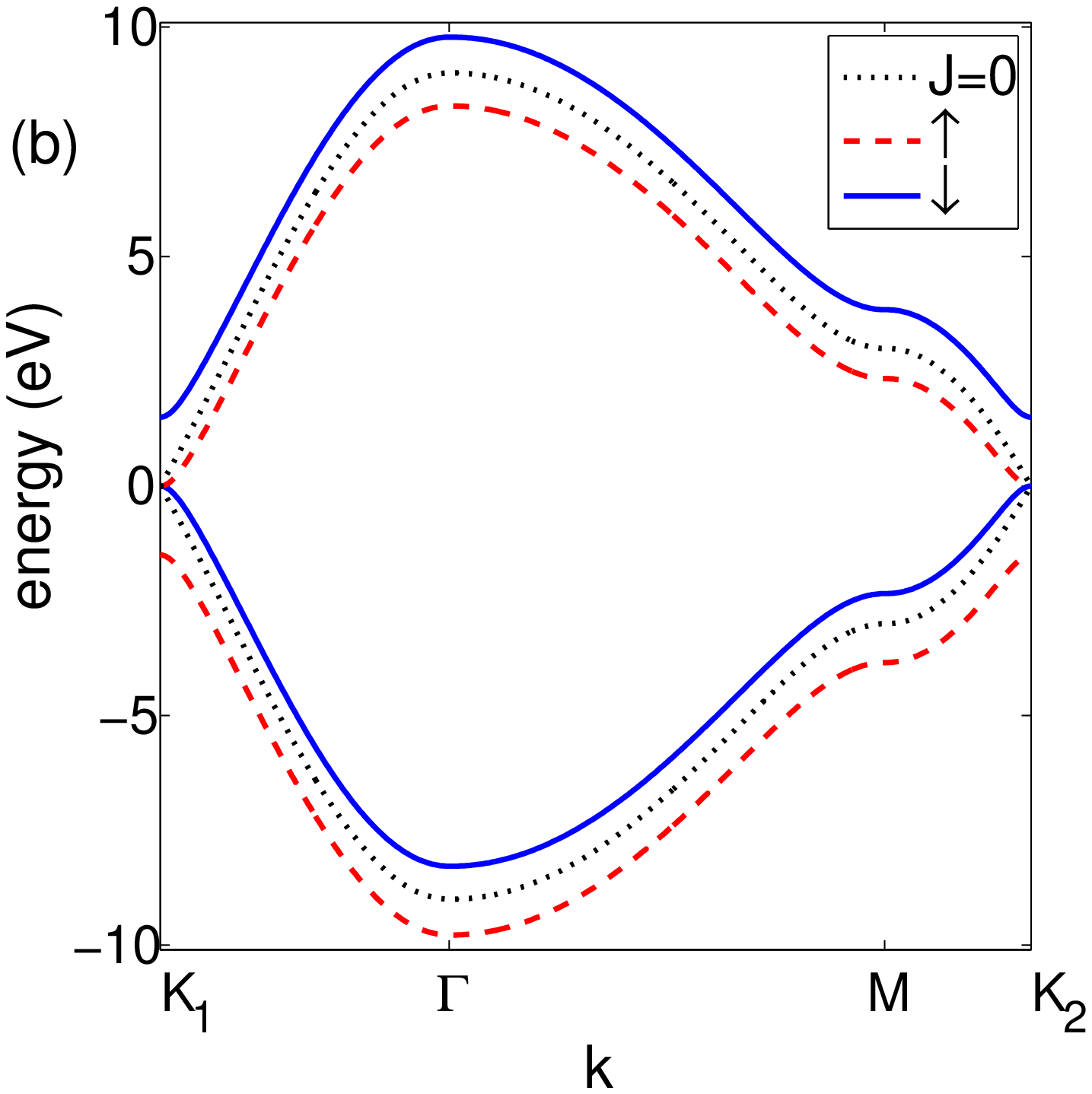}\label{fig:bands_same}}
\caption{(Color online) Dispersion for electrons that couple
  with $J=1.5\;\textrm{eV}$ to perfectly ordered MI spins. (a)
  Eq.~(\ref{eq:bands_af}) for AF ordered MIs on all sites. The curves for up and down spins lie on top of each
  other. (b) Eq.~(\ref{eq:bands_fm}) for MIs on one sublattice only;
  all impurity spins (IS) are parallel and `up'. The dotted line in each
  panel gives the bands for $J=0$. The inset in (a) shows the path through the first Brillouin zone.
  \label{fig:bands_af}}
\end{figure}

For $J=0$, 
the Hamiltonian can be easily diagonalized and the band structure along the path  
$K_1$-$\Gamma$-$M$-$K_2$ 
is shown as a dotted line in Fig.~\ref{fig:bands_af}.
The two C atoms of the basis lead to 
two states per momentum ${\bf k}$, with
energies 
\begin{equation}\label{eq:ham0}
\pm \epsilon({\bf k}) = \pm \sqrt{1+4\cos\frac{k_y}{2} \cos (\sqrt{3} k_x/2) + 4\cos^2\frac{k_y}{2}}\;.
\end{equation}
The momenta ${\bf k} = (k_x,k_y)$ allowed for 
$N_x \times N_y$ 
are
$k_x=4\pi n_x/(N_xa\sqrt{3})$ with  
$n_x=0,\pm 1,\dots,N_x/2$, and $k_y=4\pi n_y/N_y $ with 
$n_y=0,\pm 1,\dots,N_y/3$,\cite{ceulemans} and  the two 
bands are degenerate at the points $K_1$ and $K_2$.\cite{Dress}

\begin{figure}
\subfigure{\includegraphics[width = 0.4\textwidth, trim = 20 50
    10 80, clip]
{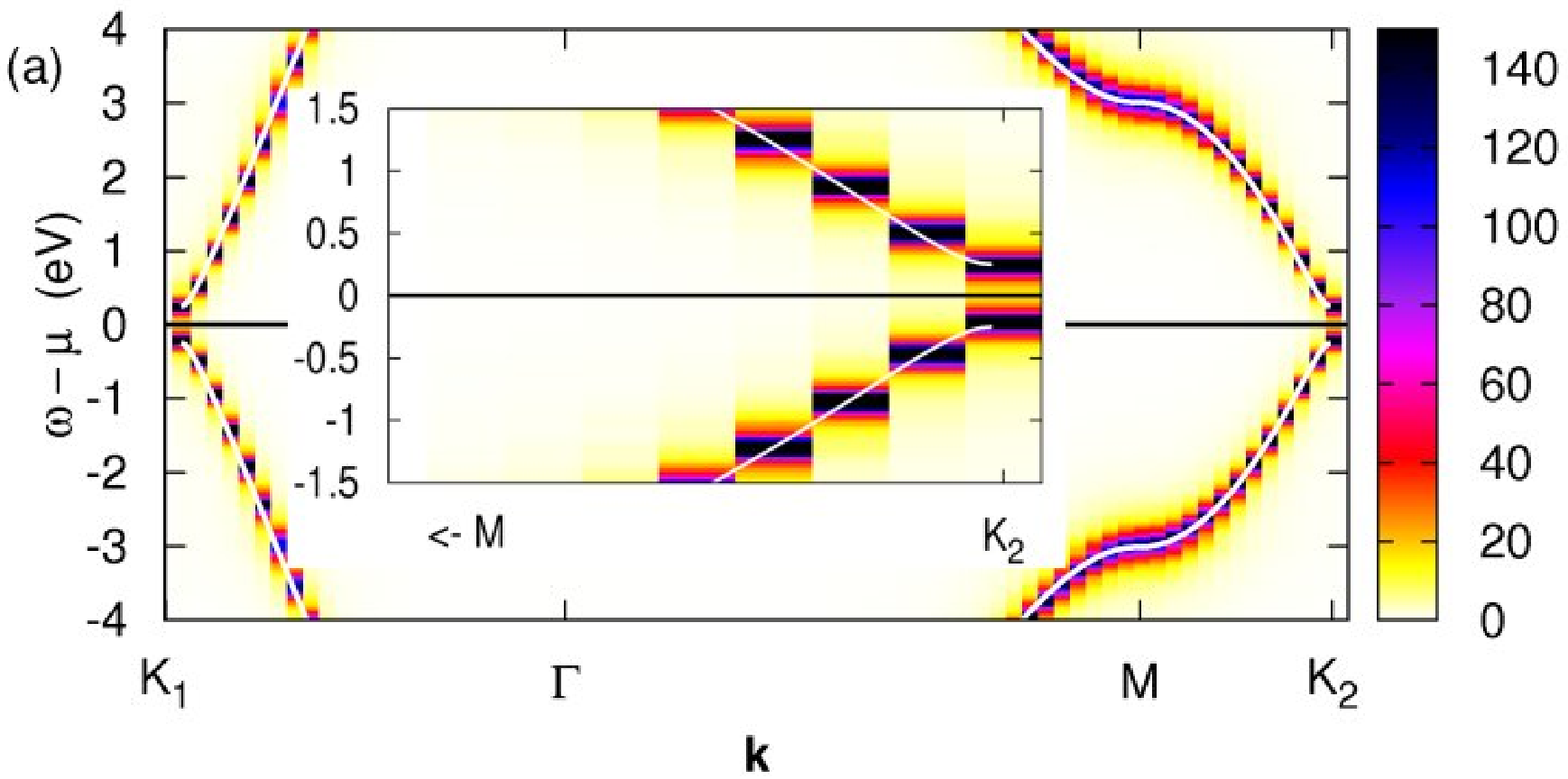}\label{fig:Ak_L72_both}}\\[-1.5em]
\subfigure{\includegraphics[width = 0.4\textwidth, trim = 20 50
    10 80, clip]
{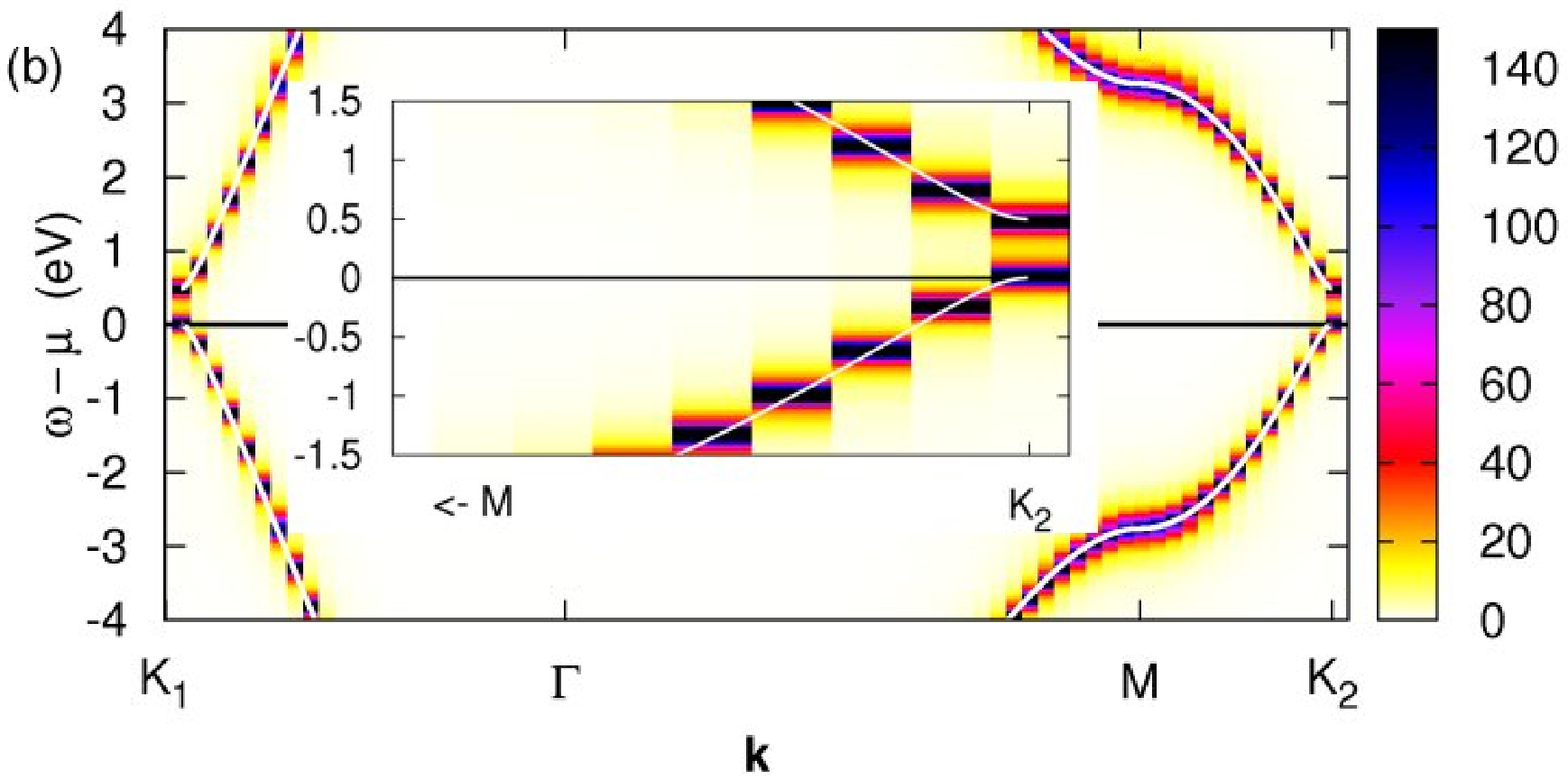}\label{fig:Ak_L72_same}}
\caption{(Color online) Spectral density for down electrons. (a)
  IS are in both sublattices; the spectral
  density for up electrons looks identical. (b) IS are in
  sublattice $A$ only and oriented along$+{\bf z}$, i.e., the down
  electrons are anti-parallel. $A({\bf k},\omega)$ for up electrons is
  the mirror image about the chemical potential $\mu$, see
  Fig.~\ref{fig:bands_same}. The white lines indicate the bands of 
  translationally invariant systems described by (a)
  Eq.~(\ref{eq:bands_af}) and a renormalized $\hat J = c_{\textrm{imp}}J$
  and (b) Eq.~(\ref{eq:bands_fm}) and $\hat J = 2c_{\textrm{imp}}J$.
  Parameters are: $N_x\times N_y = 72\times
  72$, $864=72^2/6$ impurities, $J=1.5\;\textrm{eV} = t/2$. \label{fig:spec_t0}}
\end{figure}

We can also diagonalize Eq.~(\ref{eq:ham}) for one MI per 
lattice site with all the MIs in sublattice $A$ ($B$) parallel 
 to $\hat{\bf z}$ ($-\hat{\bf z}$).
The  spin-degenerate bands are given by
\begin{equation}\label{eq:bands_af}
  E({\bf k}) =  \pm \sqrt{J^2 + \epsilon({\bf k})^2}\;.
\end{equation}
Since  \emph{non-}magnetic impurities need to be concentrated on one
sublattice to induce a gap, the resulting gap of size $2J$ [see
Fig.~\ref{fig:bands_both}] suggests that 
MIs might provide a simpler way to turn graphene into a semiconductor.

A different situation arises if all MI's are in one 
sublattice and parallel to $\hat{\bf
  z}$. The energy then depends on the 
spin and for the spin parallel (antiparallel) to $\hat{\bf
  z}$, denoted by `up' (`down'),  we find
\begin{equation}\label{eq:bands_fm}
  E^\uparrow({\bf k}) = -\frac{J}{2} \pm \sqrt{\frac{J^2}{4} + \epsilon({\bf
  k})^2},\quad
  E^\downarrow({\bf k}) = \frac{J}{2} \pm \sqrt{\frac{J^2}{4} + \epsilon({\bf
  k})^2}\;.
\end{equation}

The resulting bands are shown in
Fig.~\ref{fig:bands_same} and not surprisingly, the electrons
parallel to the localized spins are energetically
favored. But as one band for each spin direction lies 
below the chemical potential, there is no net
polarization of the itinerant electrons despite the fact
that all localized spins are perfectly FM. This follows from
the degeneracy induced by the two-atom basis and it means that
the electrons in the
``clean'' sublattice are polarized in the \emph{opposite}
direction so that the contributions from the electrons in the two sublattices 
cancel exactly. 
While there is thus no ferrimagnetism and no net spin polarization of
the itinerant electrons, the band structure in
Fig.~\ref{fig:bands_same} nevertheless shows that there are
different energy regions with well developed majority spins: in
particular, the electrons \emph{directly below and above the Fermi level are
polarized}.

\begin{figure}
\subfigure{\includegraphics[width =0.23\textwidth]
{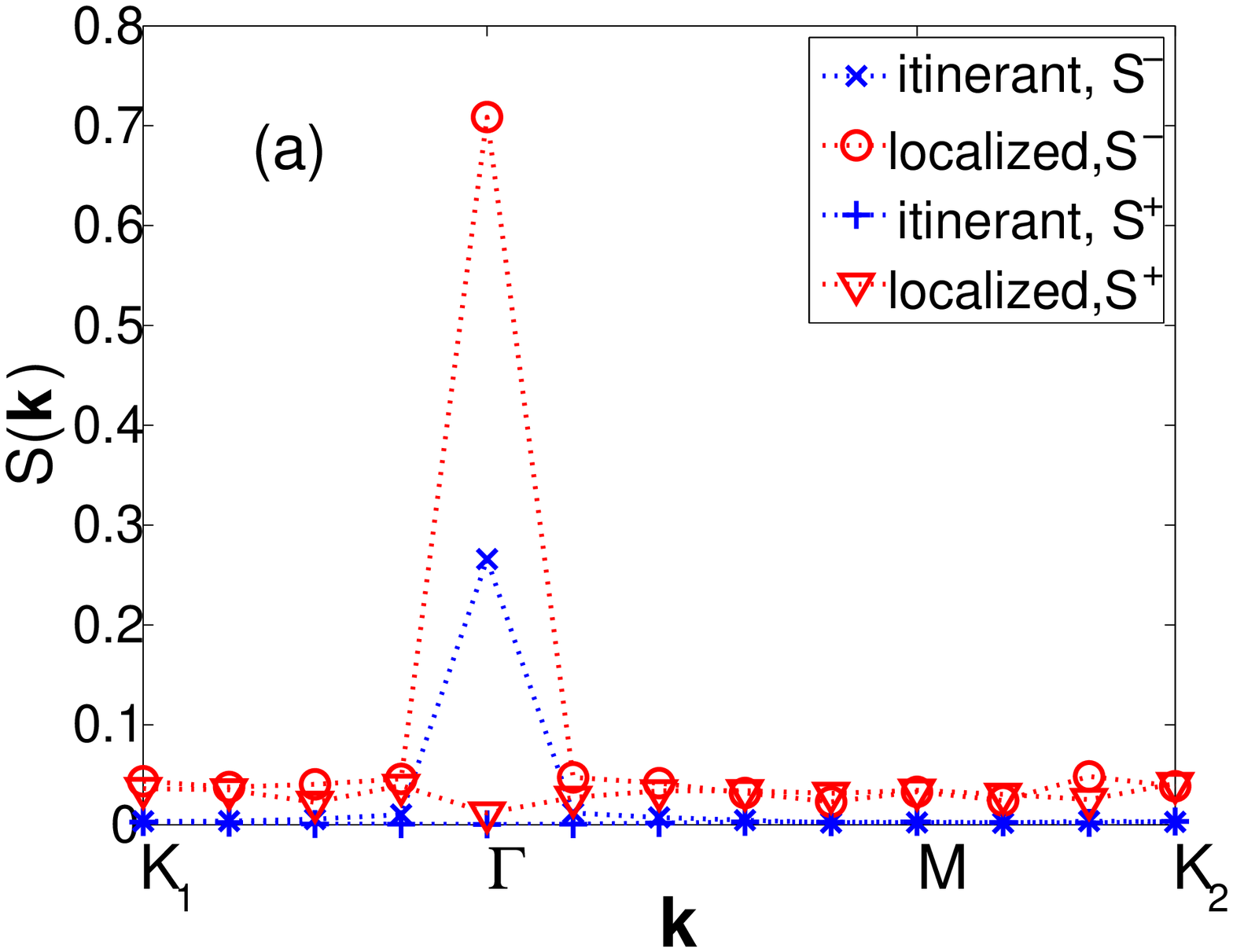}\label{fig:sk_J2_half}}
\subfigure{\includegraphics[width =0.23\textwidth]
{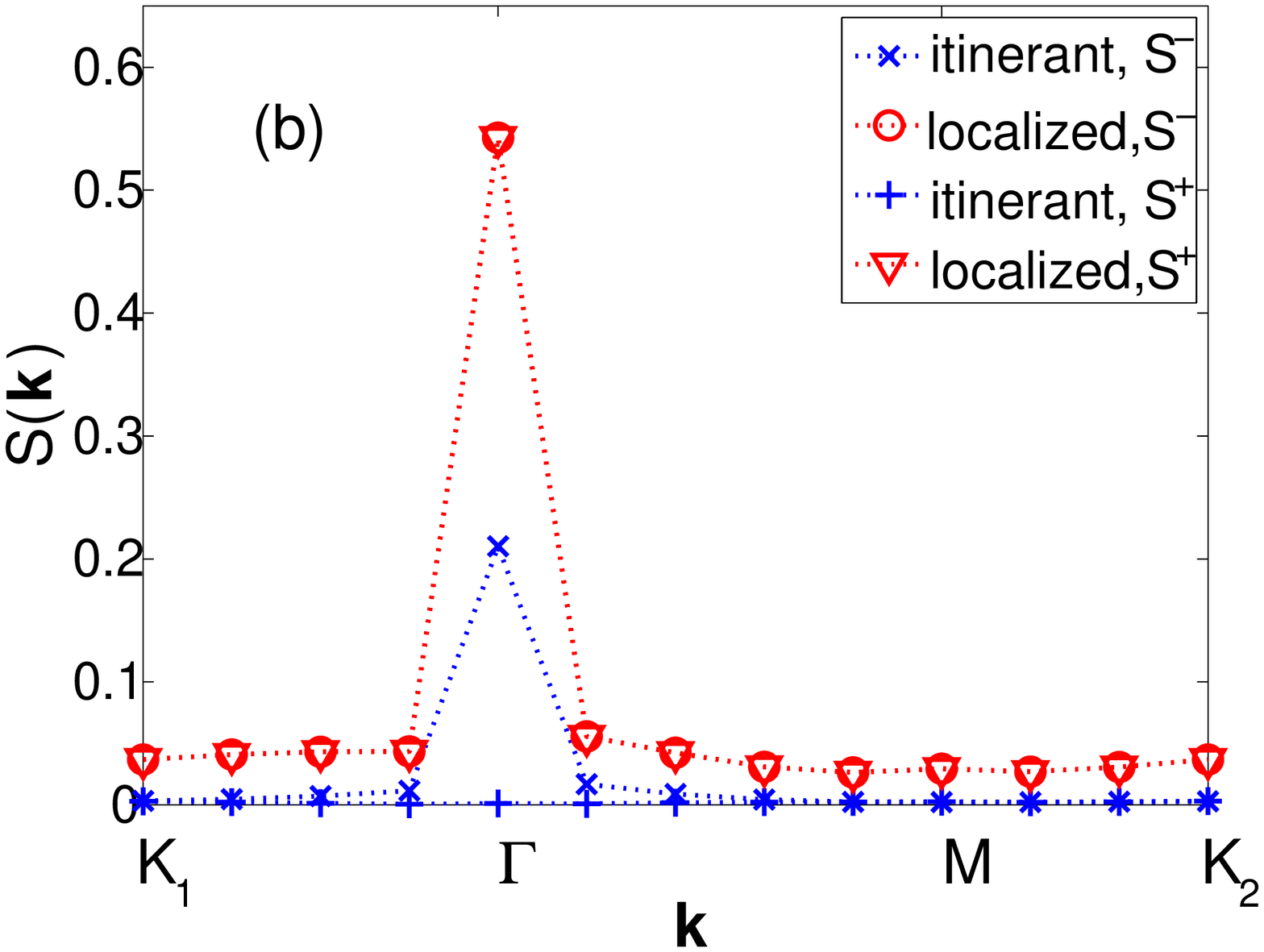}\label{fig:sk_J2_odd}}
\caption{(Color online) Spin structure factor 
  for momenta along the path shown in the inset of
  Fig.~\ref{fig:bands_both}. Symmetric ($S^+$) and antisymmetric ($S^-$)
   components indicating FM and AF order are shown for the localized IS
   as well as for the itinerant electrons for (a) each sublattice
  containing half of the randomly distributed 
  MI's and (b) all the MI's randomly distributed
  in the $A$ sublattice. $J=6\;\textrm{eV}$, $\beta t=100$,
  corresponding to 
  $T\approx 300\;\textrm{K}$,
  $N_x=N_y=12$, $N_{\textrm{imp}}=24=12^2/6$. The
  data were averaged over (a) six and (b) 
  ten impurity configurations.\label{fig:sk_J2}}
\end{figure}

As long as $J$ is not too large ($J\lesssim 2t$), the results remain
similar even if not all sites of the (sub-)lattice have a 
MI, and the ordered AF leads to dispersive bands instead of localized impurity
states. To show this, the Hamiltonian given by Eq.~(\ref{eq:ham}) was
diagonalized 
numerically for lattices with $N_x\times N_y=72\times 72 = 5184$ sites 
containing
$N_{\textrm{imp}}=864 = 72^2/6$ randomly distributed MI's.
The IS were set into a magnetically
ordered state, where all spins in sublattice $A$ ($B$) are parallel 
(antiparallel) to $\hat{\bf z}$. The 
spectral function  $A({\bf k},\omega)$ is shown in
Fig.~\ref{fig:Ak_L72_both} for equal numbers of MI's in both
sublattices. (Notice that the bands for spin up and down are here
degenerate.) Similar to the case of non-magnetic 
impurities,\cite{graphene_imp_sublatt} the most prominent change is
that the gap is renormalized from $2J$ to
$2c_{\textrm{imp}}J$ for an impurity concentration $c_{\textrm{imp}} =
N_{\textrm{imp}}/(N_x N_y)$. This can be seen by comparing 
$A({\bf k},\omega)$ and the white line given by
Eq.~(\ref{eq:bands_af}) with a renormalized $\hat J =
c_{\textrm{imp}}J$. The inset shows the resulting gap of $2 \hat J =
2c_{\textrm{imp}}J = 0.5\;\textrm{eV}$.
Doping with MIs thus turns graphene into a semiconductor via the AF
ordering of both the MIs and the graphene electrons.

In Fig.~\ref{fig:Ak_L72_same}, we present the band dispersion for the `down' 
electrons (i.e., antiparallel to the IS) for 864 MIs
randomly distributed in one of the sublattices. Since the sublattice containing the MIs 
effectively has a doubled impurity concentration of $c_{\textrm{eff}}=2c_{\textrm{imp}} =
N_{\textrm{imp}}/(N_x N_y/2)$, the renormalized gap resulting from
Eq.~(\ref{eq:bands_fm}) is given by $\hat J = 2c_{\textrm{imp}}J
= 0.5\;\textrm{eV}$, the same as for impurities in both
sublattices. The gap is thus 
independent of the percentage of impurities found in
each sublattice. The inset of Fig.~\ref{fig:Ak_L72_same} reveals that the gap now opens above the 
chemical potential $\mu$, while the gap for spin `up' (not shown) is
below $\mu$. (The up and down bands are mirror images about $\mu$.)
When all MI's are in one sublattice, the highest occupied states
are thus clearly spin polarized, leading to the  
{\it spin-polarized half-semiconductor} described 
in Ref.~\onlinecite{cervantes-sodi:165427}. 

If the MI distribution is not perfect and a fraction $p<1$ of the 
MIs are in one sublattice and $1-p>0$ in the other, the exact degeneracy of up and
down states at $K_1$ and $K_2$ is lifted and a gap of order $4(1-p)Jc_{imp}$ 
opens between the highest occupied `down' states and the lowest
empty `up' states. The gaps for each spin direction bands
remain asymmetric about $\mu$ as long as $p\gg 0.5$ (or $p\ll 0.5$),
thus preserving the spin polarization for a  reduced energy window
$2(2p-1)Jc_{\textrm{imp}}$. Again, the translationally invariant and the dilute
systems behave similarly: The ordered state of the MIs preserves coherent electron
motion in highly dispersive bands (albeit with a finite mass), and we
only find localized impurity states with small dispersion for very
large $J\gtrsim 2t$.  

After discussing electrons interacting with perfectly ordered spins,
we present results of unbiased MC
simulations. They were performed on the angles $\theta_{\bf
  I}$ and $\phi_{\bf I}$ defining the orientation of 
the IS in the Hamiltonian Eq.~(\ref{eq:ham}); 
the probability of a spin configuration is obtained by
diagonalizing the effective electronic Hamiltonian.\cite{man}
Studies for lattice sizes up to 
$18 \times 18$, for 150K$<T<$ 1500K, 
for various values of $J$, and for different impurity concentrations showed
indications of long-range magnetic order. 

Each momentum ${\bf k}$ has a spin-structure factor $S_A({\bf k})$
for sublattice $A$ and $S_B({\bf k})$ for sublattice $B$. They
indicate the magnetic order within the sublattice, with a signal at
the $\Gamma$-point (${\bf k} =0$) signifying FM order between all
spins in the same sublattice. $S_A$
and $S_B$ can be combined into symmetric ($S^+ = S_A+S_B$) and
antisymmetric ($S^-= S_A-S_B$) combinations, which give
information about the order between the sublattices: If the spins in
sublattice $A$ and $B$ are parallel (antiparallel), this is revealed
by a signal in $S^+$ ($S^-$). 
In Fig.~\ref{fig:sk_J2}, $S^+$ and $S^-$ 
are shown as a function of the momentum.  
One sees clear peaks at $\Gamma$, indicating
that the itinerant electrons mediate order between the localized
spins and that all spins within a sublattice are FM. 
For MIs located in both sublattices, the signal is in the $S^-$ channel [see 
Fig.~\ref{fig:sk_J2_half}], which shows AF order between the
sublattices. 

If sublattice $B$ is free from MIs, it does not contribute to the
impurity-spin structure factor, leading to $S^+=S^-=S_A$,
and the signal at $\Gamma$ indicates FM order within the MI-doped
sublattice [see Fig.~\ref{fig:sk_J2_odd}]. The MC simulations thus
clearly demonstrate FM order within sublattices 
and AF order between them, justifying the choice of ordered
ground-state configurations discussed above.\cite{snap} The structure factor of the itinerant electrons
develops a peak at $\Gamma$ in the AF $S^-$ channel regardless of the
MI distribution. Since there is \emph{no} signal in the
symmetric channel $S^+$ (it has 
been verified that this also holds for momenta not  
on the path shown in Fig.~\ref{fig:sk_J2}), ferrimagnetism 
can be ruled out and the itinerant electrons do not develop a net
magnetic moment, as discussed previously for the perfectly
ordered IS.

For small lattices, the few available momenta mean a large finite-size
energy spacing of the kinetic energy, and an asymmetric gap, as
seen in Figs.~\ref{fig:bands_same} and~\ref{fig:Ak_L72_same} does then
not show up very clearly. 
The smaller lattices accessible to MC require thus the use of
larger values of $J \gtrsim t$ to allow us to analyze $A({\bf
k},\omega)$. 
Where the analysis is possible, 
the MC data agree with the ground-state results.
For MIs in only one sublattice, the spin polarization is
reduced by finite-temperature fluctuations, but the states around
the FS remain predominantly `down' ($\omega \lesssim \mu$)
and `up' ($\omega \gtrsim \mu$). The unbiased MC calculations thus
corroborate the results for magnetically ordered configurations. 

\section{Conclusions}\label{sec:conclusions}

We have discussed  itinerant electrons in graphene
interacting with MIs localized on top of the C atoms. We found
that the electrons always 
develop AF correlations and that a gap opens for both
spin projections, regardless of the distribution of the MI's
between the two sublattices. 
Domain boundaries or edges of
nanoribbons~\cite{cervantes-sodi:165427,graphene_imp_sublatt,Yao:2009p2427}
are expected to support edge states within the gap. 
For MI's randomly distributed between the
sublattices, 
we find a 
standard semiconductor with a gap that can be tuned by adjusting the
impurity concentration. While the carriers acquire a mass, their
dispersion remains almost that of clean graphene. If all or most
MI's are in one sublattice~\cite{foot2} 
the highest occupied
(lowest unoccupied) states 
are spin polarized antiparallel (parallel) to the MIs, giving rise to a 
spin-polarized semiconductor that could be useful to inject polarized 
electrons into devices.
We find that MIs with large spin  located on top of C 
atoms are the most likely candidates to produce these effects.
\vspace*{-2em}
\acknowledgments
\vspace*{-1em}

This work was supported by the National Science Foundation grant DMR-0706020, 
the Division of Materials Sciences and Engineering, Office of Basic Energy
Sciences, U.S. DOE, and the DFG under the Emmy-Noether program.


\begin{thebibliography}{10}
\bibitem{foot1}
Several locations for MIs have been considered in the literature. We are
  treating MI's that interact mostly with the electrons at one particular
  lattice site. This corresponds, e.g., to substitutional impurities or moments
  induced by vacancies.\cite{Yaz,Voz,Brey,Duffy}
\bibitem{foot2} This is the distribution needed
to achieve a gap with non-MI's~\cite{graphene_imp_sublatt}.
\end{thebibliography}

\end{document}